\documentclass[floatfix,reprint,amsmath,amssymb,prapplied,superscriptaddress,aps]{revtex4-2}
\usepackage[dvipsnames]{xcolor}
\usepackage{graphicx}
\usepackage{dcolumn}
\usepackage{bm}
\usepackage{lipsum, babel}
\usepackage{siunitx}
\usepackage{hyperref}
\hypersetup{
colorlinks=true, breaklinks=true, bookmarks=true,bookmarksnumbered,
urlcolor=RoyalBlue, linkcolor=RoyalBlue, citecolor=RoyalBlue, 
pdftitle={},
pdfauthor={\textcopyright}, 
pdfsubject={},
pdfkeywords={}, 
pdfcreator={pdfLaTeX},
pdfproducer={LaTeX}
}
\begin{document}

\title{Experimental realization of metastable target skyrmion states in continuous films}

\author{Elizabeth M. Jefremovas*}
\affiliation{Institute of Physics, Johannes Gutenberg University Mainz, Staudingerweg 7, 55128 Mainz, Germany}
\email{martinel@uni-mainz.de; **klaeui@uni-mainz.de}
\author{Noah Kent}%
\affiliation{Research Laboratory of Electronics, Massachusetts Institute of Technology, Cambridge, MA, United States of America}%
\author{Jorge Marqu\'es--March\'an}
\affiliation{Institute of Materials Sciences of Madrid -- CSIC, 28049 Madrid, Spain}
\author{Miriam G. Fischer}
\affiliation{Institute of Physics, Johannes Gutenberg University Mainz, Staudingerweg 7, 55128 Mainz, Germany}
\author{Agustina Asenjo}
\affiliation{Institute of Materials Sciences of Madrid -- CSIC, 28049 Madrid, Spain}
\author{Mathias Kl{\"a}ui**}
\affiliation{Institute of Physics, Johannes Gutenberg University Mainz, Staudingerweg 7, 55128 Mainz, Germany}
\email{klaeui@uni-mainz.de}

\date{\today}

\begin{abstract}
Target skyrmions (TSks) are topological spin textures where the out-of-plane component of the magnetization twists an integer number of $m$-$\pi$ rotations. Based on a magnetic multilayer stack in the form of $n\times$[CoFeB/MgO/Ta], engineered to host topological spin textures via dipole and DMI energies, we have successfully stabilized 1$\pi$, 2$\pi$ and 3$\pi$ target skyrmions by tuning material properties and thermal excitations close to room temperature. The nucleated textures, imaged via Kerr and Magnetic Force Microscopies, are stable at zero magnetic field and robust within a range of temperatures (tens of Kelvin) close to room temperature (RT = 292 K) and over long time scales (months). Under applied field (mT), the TSks collapse into the central skyrmion core, which resists against higher magnetic fields ($\approx$ 2 $\times$ TSk annihilation field), as the core is topologically protected. Micromagnetic simulations support our experimental findings, showing no TSk nucleation at 0 K, but a $\approx$ 30 $\%$ probability at 300 K for the experimental sample parameters. Our work provides a simple method to tailor spin textures in continuous films, enabling free movement in 2D space, creating a platform transferable to technological applications where the dynamics of the topological textures can be exploited beyond geometrical confinements.
\end{abstract}

\maketitle

\begin{figure}
\includegraphics{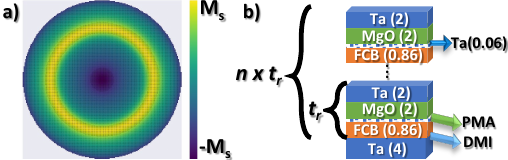}
\caption{\label{stack} \textbf{a)} Top-view of the magnetization of a $2\pi$ target skyrmion, also known as \textit{skyrmionium}. Scale colour indicates the direction of the $z$ component or the magnetization $M$, normalized to $M_{s}$. \textbf{b)} Scheme of the multilayer stack. Each multilayer unit, $t_{r}$, is composed by (from bottom to top) Ta, FCB (Fe\textsubscript{60}Co\textsubscript{20}B\textsubscript{20}), MgO and Ta, number in brackets referring to the film thickness (in nm). Each unit is repeated $n$ times to scale up the dipolar coupling. All samples are deposited on Si/SiO\textsubscript{2}.}
\end{figure}

Magnetic skyrmions are topological spin solitons with potential applications in spintronic devices, owing to their size, mobility and topological stabilization \cite{liu2016skyrmions, fert2017magnetic, finocchio2016magnetic, everschor2018perspective}. Along with skyrmions, further chiral magnetic topological spin textures have gathered the interest from the scientific community, such as merons, chiral bubbles or target skyrmions (TSks) \cite{wang2021meron, zheng2017direct, cortes2019nanoscale}. The latter $m$--$\pi$ vortices, experimentally demonstrated in the last decade, differ from skyrmions by the higher magnetization twists along the $z$ component, quantified by $m>1$ [see Fig.~\ref{stack}\textbf{a}]. Beyond their fundamental interest, an even number of rotations $m$ leads to a trivial topological charge, as defined by the winding number $\mathcal{W} =\frac{1}{4\pi} \int dxdy\; \vec{m} \; (\partial_{x}\vec{m}\;\partial_{y}\vec{m} )$. In the absence of topological charge, there is no skyrmion Hall effect, which is advantageous for current--driven technology applications \cite{komineas2015skyrmion, li2018dynamics, zhang2016control, tang2021magnetic, gobel2019electrical, litzius2017skyrmion}. In addition to this, the polarity of a TSk can be easily switched by means of AC field, which provides an additional control parameter in spintronics applications \cite{vigo2020switching}. \newline

In the recent years, TSks have also enabled the experimental realization of 3D spin textures, challenging to be stabilized in experiments, as they serve as \textit{hopfion precursors} \cite{kent2021creation}. These topological 3D solitons, together with other recently discovered structures like skyrmion cocoons \cite{grelier2022three}, are the result of the expansion into the 3rd dimension of 2D topological textures, a groundbreaking research field hosting exciting novel physical phenomena and potential applications in magnetic sensor and information processing technologies \cite{ran2021creation, grelier2022three, fischer2020launching, kent2021creation, fernandez2017three}. There is, however, an open challenge in the experimental stabilization of 3D solitons for technological applications. Up to date, the preferred strategy relies on the lateral confinement of the structures (typically, in a nanodisk shape). This procedure, although successful for stabilizing hopfions \cite{kent2021creation}, actually prevents the free motion of the soliton in the whole continuous film, hampering their foreseen computing applications, as the information carriers need to freely move in space \cite{markovic2020physics, zhang2023magnon}. \newline

Here, we report a magnetic multilayer stack, capable of hosting target skyrmions with and without finite topological charge (1$\pi$ and 3$\pi$, and 2$\pi$-type) in the continuous film. The TSks are activated through temperature fluctuations close to room temperature (RT = 292 K), being stable in a range of temperatures close to RT and zero field. In addition to this, we report an increase of the energy barrier of such metastable states with the magneto-dipolar coupling strength, scaled up by increasing the number of repetitions of the multilayer ($n =$ 2, 3, 4, 7, 12). To ensure the technological transfer of our results, the multilayer stack is designed so as the metastable TSk states can be reached at temperature values in agreement with the working range of technological devices (\textit{i.e.}, from 314 K up to 380 K), which validates our method for application perspectives. \newline

We have designed a magnetic multilayer stack in the form of Ta(4)/[Co\textsubscript{20}FeCo\textsubscript{60}B\textsubscript{20}(0.86)/MgO(2)/Ta(2)]\textsubscript{n}, being $n=$ 2, 3, 4, 7 and 12. Numbers in brackets indicate the layer thickness, in nm. The Ta/CoFeB interface provides the interfacial Dzyaloshinskii-Moriya interaction (DMI), while the CoFeB/MgO interface is the source of the perpendicular magnetic anisotropy (PMA) of the stack. Note the selection of MgO, an insulator material, which ensures that the only type of interlayer coupling is of dipolar magneto-static origin. This can be gradually scaled up by increasing the amount of magnetic material (\textit{i.e.} the number of repetitions $n$). We have also introduced a Ta dusting layer (nominal thickness below 1 monolayer), a strategy that reduces the PMA as well as minimizing the pinning of the stack \cite{gruber2022skyrmion, gruber2023300, zazvorka2019thermal, eli}. A schematic representation of the stack can be visualized in Fig.~\ref{stack}\textbf{b}. To image the TSks, we have employed Kerr Microscopy in polar mode, and Magnetic Force Microscopy (MFM) in Variable--Field (VF--MFM) mode, which allows us to probe structures down to sub--nm regime while applying an OOP magnetic field. Details on the particular set up can be found in Ref. \cite{jaafar2009variable}. All the stacks were heated by a Peltier element equipped with a Pt100 sensor attached right next to the sample for temperature control between 280 and 380 K. The temperature stability was ensured to be within 0.1 K from 280 to 350 K, and within 1 K up to 380 K. \newline

\begin{figure*}
\resizebox{2.0\columnwidth}{!}{\includegraphics{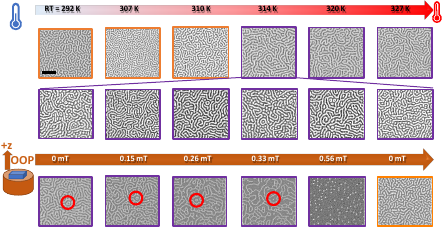}}
\caption{\label{temp_cycles} Kerr Microscopy images of the TSks stabilized in a 200 $\mu$m square patterned for the $n =$ 2 stack. Scale bar 10 $\mu$m is common for all images, black (white) contrast represents magnetization in $+z$ ($-z$) direction. All measurements are performed at RT = 292 K. Top (middle) panel evaluates the thermal nucleation (stability) of TSks at $\mu_{0}H =$ 0 mT OOP. Bottom panel evaluates the stability of the TSks under out of plane (OOP) magnetic field applied in $+z$ direction. Multidomain patterns (TSk) are enclosed in orange (purple). In bottom image, the annihilation of a TSk can be followed by the red circle.}
\end{figure*}

Fig.~\ref{temp_cycles} provides an overview on the nucleation procedure of the TSk states in $n = 2$. To ease the tracking of the nucleated textures, while ensuring the reproducibility of the experiments, a 200 $\mu$m square has been patterned using Electron Beam Lithography. The large dimensions of the structure (200 $\mu$m) exclude significant lateral confinement effects that may yield differences to the continuous film behaviour (see Supplemental Material, Fig.~S1). This strategy has enabled us to repeat systematically the described nucleation procedure excluding thermal drift or other error sources, confirming its validity and robustness against time. \newline

\begin{figure}\resizebox{1.0\columnwidth}{!}{
\includegraphics{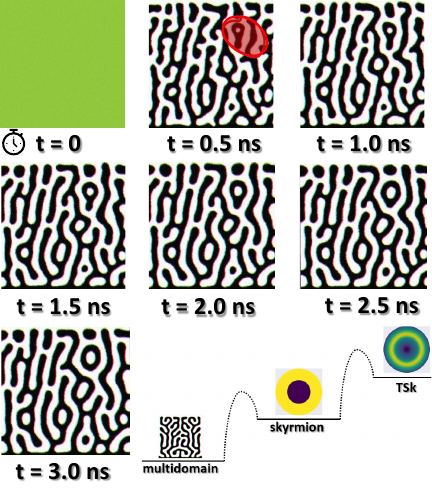}}
\caption{\label{simulations_2} Time-lapse of the nucleation of a target skymion according to nanomagnetic simulations, 0.5 ns snapshots. Black and white colors represent magnetization in $+-z$ direction, respectively. Encircled, we see the formation of a skyrmion, which is further surrounded by a magnetic domain. After 3.0 ns, the structure is closed and forms a TSk. Bottom right a schematic showing relative energies of stable and metastable states in this system, from multidomain to TSk through skyrmion  }
\end{figure}

The ground state of the system after saturation at RT = 292 K corresponds to a multidomain pattern, as it can be seen in the top left image, encircled in orange (Fig.~\ref{temp_cycles}). In order to activate TSk states using temperature fluctuations, the following nucleation procedure is followed: after IP saturation, the sample temperature is stabilized at the values inserted at top row of Fig. ~\ref{temp_cycles} (307, 310, 314 K...). Then, the temperature is decreased back to RT. As it can be observed in the top row of Fig.~\ref{temp_cycles}, below $T =$ 314 K, the sample relaxes back to a multidomain configuration without significant alteration of the domain periodicity (see images enclosed in orange in Fig.~\ref{temp_cycles}). However, the situation changes drastically for $T \geq$ 314 K ($\approx $ 27 meV), when $m-\pi$ TSk states get activated, including $m =\;$1$\pi$ (skyrmions), 2$\pi$ (also known as \textit{skyrmionium}) and even, 3$\pi$. Further increasing the thermal excitation to 320 or 327 K also yields TSks (see images enclosed in purple in Fig.~\ref{temp_cycles}). It is worth noting that the number of TSks is on average the same, regardless the temperature value, indicating that the probability to nucleate TSk remains constant in the whole studied temperature range ($\leq$ 327 K). This reveals that the energy landscape of our stack hosts metastable topological phases, protected by energy barriers that can be overcame at temperature ranges close to RT (see below Fig.~\ref{simulations_2}). In addition to this, the TSks are nucleated at different positions in every re-nucleation for all the stacks ($n =$ 2, 3, 4, 7, 12). Therefore, there is no preference of some specific sites for the nucleation of the textures. This indicates a flat energy landscape of the sample, thus pinning is not the mechanism responsible for TSk nucleation, which constitutes a step forward compared to previous works \cite{zheng2017direct, kent2019generation}. Furthermore, our stack is also capable of stabilizing TSks following the previously reported strategy based on  the geometrical confinement. As it can be seen in Fig.~S1 in Supplemental Material, TSks are stabilized at zero field in 4 $\mu$m microdiscs, while the larger disks showcase a multidomain state with a similar domain periodicity of the continuous film ($\approx$ 2 $\mu$m). This means that our stack can not only host TSk in confined geometries, yet it does in the continuous film, with boosts their implementation in technology applications. \newline

With thermal nucleation of TSks achieved, we have tested the TSk stability against further thermal excitations. The middle panel of Fig.~\ref{temp_cycles} provides a comprehensive overview on the effects of temperature for the TSks nucleated at 314 K. Starting at RT, we have raised and stabilized the temperature to the lowest value (307 K), and then, decreased back to RT, where the measurements are performed, all at 0 field. Then, the temperature has been raised up to the next value (314 K), and the process is repeated until 327 K. As it can be seen in the whole series, the TSks are robust against temperature, as no significant changes are detected. We have also reversed the procedure, \textit{i.e.}, starting the temperature cycling from highest $T$ (327 K), and repeating the process reaching 320 K, 314 K... until the lowest 307 K, obtaining the same results. \newline

We have further studied the stability of the TSks against out--of--plane (OOP) magnetic field. Our experimental results, included in the bottom panel of Fig.~\ref{temp_cycles}, illustrate the stability of the TSk up to 0.33 mT, where they collapse, but with a clear difference between the annihilation of even and odd $m$--$\pi$ TSks. Fundamentally, there are only two types of $m--\pi$ skyrmions, $m =$ 1 ($\mathcal{W} = 1$), and $m =$ 2 ($\mathcal{W} = 0$). All other TSks are built up as a combination of the two. This means that a 3$\pi$ TSk is actually a 1$\pi$ TSk wrapped by a 2$\pi$ TSk, with no net topological charge. This implies a different collapse path for the structures, being the domains annihilated at lower fields than the inner skyrmion, which holds topological charge. As the OOP field is increased, the external 2$\pi$ TSk shrinks progressively until it gets annihilated at 0.33 mT (follow red circle in Fig.~\ref{temp_cycles}), in a similar manner than the 2$\pi$ domain wall collapses in the 1 dimensional case \cite{kent2019generation, cortes2017thermal, rohart2016path, zhang2018real}. On the contrary, the central 1$\pi$ TSk remains, since its collapse requires going through a Bloch point, implying a higher energy to be annihilated \cite{rohart2016path}. On the other hand, for the case of the odd--$\pi$ TSks, their topological charge is $\mathcal{W}$ = 1, meaning that an additional (topological) energy barrier needs to be overcome to annihilate \cite{ognev2020magnetic}. For this, all collapse paths require a transition through an energetically highly unfavorable magnetic state, resulting in a skyrmion surrounded by a chiral horseshoe domain \cite{kent2019generation}. This additional protection explains why at 0.56 mT, odd--$\pi$ TSk still survive, while the even--ones have already collapsed. It is worth mentioning here the recent work by Nakamura \textit{et al.} \cite{nakamura2024communicating}, where they propose to adopt the perspective of \textit{communication between skyrmions}. Adopting their perspective, the annihilation path can be viewed as the energy competition between the opposite polarities (skyrmion core \textit{vs.} ring). Higher rotations of the magnetization ($m$) would result into increased competition of polarities, ensuring a more robust TSk. \newline

Finally, we have tested the validity and robustness of our thermal-excitation TSk nucleation method against magneto-dipolar interactions. This constitutes an essential step to assess the validity of our proposed experimental methodology in building up a 3D spin texture, following the procedure stated by N. Kent \textit{et al.} \cite{kent2019generation}, where a 3D hopfion texture can be promoted by inserting a TSk in between two multilayer stacks with strong PMA, forcing the magnetization to wrap into the whole space. Important to note, the current experimental efforts for stabilizing 3D solitons notably rely on the expansion of the magnetic multilayer stack into the $z$-direction, as the multilayer stack allows to tune exchange, DMI, anisotropy or dipolar interactions. However, as a consequence of the increasing number of repetitions, the importance of the dipolar magnetostatic coupling raises, which could imply enhanced energy barriers protecting the TSk energy states. This would shift the thermal energy up, with the inherent thermal fluctuations, that could potentially prevent the stabilization of any spin texture that may be formed. In order to address this critical issue, we have reproduced the thermal-excitation nucleation procedure in 4 stacks with increasing number of repetitions, from $n =$ 3 to $n =$ 12, keeping the same composition as the one described before for $n =$ 2 [Ta/CoFeB/MgO/Ta]. \newline

\begin{figure}\resizebox{1.0\columnwidth}{!}{\includegraphics{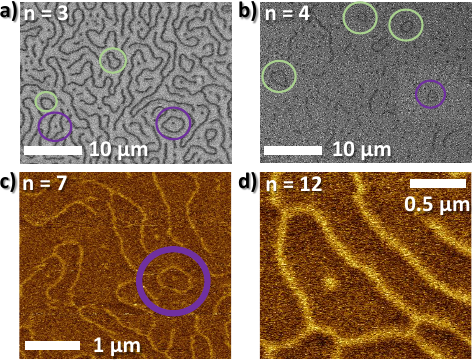}}
\caption{\label{multilayers} Room temperature measurements of $n=$ \textbf{a)} 3, \textbf{b)} 4, \textbf{c)} 7 and \textbf{d)} 12 repetitions, nucleated at OOP field of $\mu_0{H} =$ 0.73, 6.3, 11.8, and 26.1 mT, respectively. Green circles enclose some of the 2$\pi$ TSk, while the purple ones, the 3$\pi$ TSk.}
\end{figure}

Fig.~\ref{multilayers} includes representative images of TSk generated thermally in $n =\;$3, 4, 7 and 12 repetitions. After saturation at room temperature, all the continuous films showed a multidomain maze-like pattern state. To nucleate the TSk, an OOP field was needed, unlike the case of $n=\;2$, along with a higher temperature ramp. In such a way, we managed to nucleate TSk by an OOP field of $\mu_0{H} =$ 0.73$\pm$0.01, 6.30$\pm$0.01, 11.8$\pm$0.1, and 26.1$\pm$0.1 mT in $n=\;$ 3, 4, 7 and 12, at temperature values of 320, 327, 350 and 380 K, respectively. The systematic increase of the nucleation temperature with increasing number of repetitions (\textit{i.e.}, dipolar interactions) indicates a higher energy barrier for the annihilation of the topological states, which agrees with magneto--dipolar coupling stabilizing role of spin textures \cite{fert2017magnetic, eli, lucassen2019tuning, kent2021creation}. Our experimental evidence points to the energy barrier to scale linearly with the number of repetition $n$ as $E \propto 7.15 k_{B}n$, on first approximation. Assuming that the only sample parameter that changes from one stack to the other is the dipolar coupling \cite{eli}, the linear dependence with $n$ comes simply as a consequence of the increase of magnetic material, as already stated in the literature, \textit{e.g.} Woo \textit{et al.} \cite{woo2016observation}. It is worth mentioning that all the experiments are performed at temperature values far below the annealing point promoting structural changes ($\approx$ 520 K or even, higher \cite{wang2006interfacial}), and in no case did the magnetic properties of the stack irreversibly change, as the multidomain state was always settled after saturation.\newline

We would like to highlight the existence not only of 2$\pi$ (see green circles in Fig. \ref{multilayers}), but also 3$\pi$--TSk states (purple circles), in our multilayer stacks, yet their occurrence decreases as the number of $m$ rotations increases. On average, the amount of 3$\pi$ TSks is half the one of 2$\pi$-ones in $n=$ 3, dropping to a fourth in $n =$ 4. Finally, it is worth mentioning the stability of the nucleated TSk over time. As an example, in Fig.~S3 in Supplemental Material, we have included an image corresponding to the TSk nucleated in $n =$ 7, which remains unaltered in the sample 4 months after having being nucleated, all preserved at zero field and room temperature. \newline

\begin{figure*}\resizebox{2.0\columnwidth}{!}{
\includegraphics{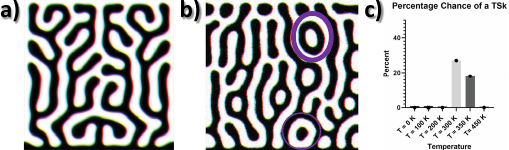}}
\caption{\label{simulations} MuMax3 Thermal Micromagnetic of the relaxed state of the system with an anisotropy of 340 kJ/m$^3$. Black and white represent spins in $+z$ and $-z$ direction, respectively \textbf{a)} $T$ = 0 K, no TSks nor skyrmions observed in any simulation. \textbf{b)} $T$ = 300 K. Two target skyrmions (encircled in purple) are stabilized. \textbf{c)} Percentage of TSks nucleated at $T =$ 0, 100, 200 (none), 300 K (maximum), 350 K and 450 K (none). 11 simulations performed at each temperature.}
\end{figure*}

We have performed micromagnetic MuMax3 simulations \cite{vansteenkiste2014design} in order to understand the role thermal fluctuations in the nucleation of TSk states. For these, we have used a 2 $\mu$m $\times$ 2 $\mu$m $\times$ 24 nm slice of material with similar magnetic properties as the CoFeB multilayer stacks at $n =$4 ($M_{s}$ = 600 kA/m, $K_{u} = $ 340 kJ/m\textsuperscript{3}, $D = 0.9$ mJ/m$^{2}$ and $A = 10$ pJ/m, see \cite{eli}). The system was initialized to match experimental conditions (in plane saturation, followed by an OOP field of 6 mT), and allowed to relax under the external field. We slightly varied the anisotropy between simulations to account for non-uniformity in material parameters, and focused on the time scale between nucleation and stabilization, which is in the order of tens of nanoseconds. Very important to mention, none of our simulations has included an energy contribution associated to non-flat energy landscapes, excluding thus this parameter from the TSk nucleation. Six sets of simulations were performed for all material parameters: at $T$ = 0, 100, 200, 300, 350 and 450 K. Our simulation results reveal the critical role of thermal fluctuations when the system relaxes from saturation, as no target skyrmion is created at $T$ = 0 K. In contrast, the system relaxes to a multidomain pattern regardless the anisotropy value, as shown in Fig.~\ref{simulations}\textbf{a}, ensuring that the energy minima of the system is not the skyrmion lattice, \textit{i.e.}, topological states are metastable states in our system. On the other hand side, in the presence of thermal fluctuations at $T$ = 300 K, up to two TSks are nucleated and stabilized when the sample relaxes from saturation, as shown Fig.~\ref{simulations}\textbf{b}. We have further investigated the subtleties of this thermally activated process by analyzing the results of temperature values below and above 300 K. Fig.~\ref{simulations}\textbf{c)} depicts the percentage of TSk nucleated after saturation at the different simulated temperature values. Below 300 K, thermal fluctuations are not sufficient to overcome the TSk energy barrier, being the nucleation probability also zero at $T$ = 100 and 200 K. At $T$ = 300 K, the chance to get a TSk raises to 30$\%$. Further increase of the temperature yields a cut of the probability to 20$\%$ at 350 K, dropping to zero at 450 K. This reveals the fine balance of thermal fluctuations during the relaxation of the system. On the one hand, they need to deliver enough energy to the system to overcome the energy barrier protecting the metastable state, on the other hand, they must keep moderated enough to let the nucleated TSks stabilize. \newline

Worth to mention, once nucleated, the TSks are stable and robust against temperature reduction, assuming no critical temperature dependence of the magnetic properties. Micromagnetic simulations show the prevalence of the TSks down to 0 K, revealing that temperature fluctuations stabilize the spin texture as the structure relaxes after saturation. This finding matches our experimental observations, as the nucleated TSks at 350 K in $n=\;7$ where stable at 300 K and 0 field 4 months after being nucleated (see Fig.~S3 in Supplemental Material). \newline

To elucidate the role of thermal fluctuations in the nucleation process of the TSk metastable states, we have performed micromagnetic simulations focusing at very short time steps after nucleation from saturation. Fig.~\ref{simulations_2} includes the first 7 frames in time steps of 0.5 ns. We have identified that the starting point of the TSk nucleation is intimately related to the nucleation of skyrmions. As it can be seen, right immediately after saturation and heated up to 300 K, a skyrmion is nucleated (see $t$ = 0.5 ns, encircled in red), as a consequence of the reduction of the PMA with the increase of the sample temperature, which renders topological metastable states accessible. Importantly, the process does not stop after the skyrmions are activated, but thermal fluctuations further deform the magnetic domains. As a result, a magnetic domain can encircle a skyrmion, (follow red circle in Fig.~\ref{simulations_2}), which further stabilizes into a TSk provided the thermal fluctuations are sufficient to overcome the domain wall repulsion and not too high to settle down a closed TSk, as it can be seen after 3.0 ns. A sketch of the process, starting from multidomain, ending up in TSk via skyrmion metastable state is captured in the bottom right sketch of this Fig.~\ref{simulations_2}. Further increase of temperature provides higher thermal fluctuations, easing the path to overcome domain wall repulsion, in such a way as 3$\pi$ TSks, or further topological states, can be accessed. However, at much higher temperatures, thermal fluctuations will overwhelm the skyrmion like core of the TSk, causing it to collapse, leaving just a normal skyrmion. \newline

Starting from a magnetic multilayer stack, we have managed to nucleate TSk textures assisted by temperature fluctuations, with different topological charges in continuous films. The required temperature values are compliant with the ones of operating technological devices, and the described method is also valid for multilayer stacks comprising several repetitions in $z$ direction. The created 1, 2 and 3$\pi$ TSks are robust against temperature variations, low magnetic fields, dipolar coupling and time (months). Micromagnetic simulations on a material with similar properties as our stack support our results, showing a negligible probability of TSk nucleated at $T =$ 0K, in contrast to the 30$\%$ chance of getting them at 300 K. \newline

The crucial difference between our experimental system and previous ones, is that our nucleation method relies on no pinning or lateral confinement stabilizing TSk states, where all previous works did \cite{zheng2017direct, kent2019generation, zhang2016control, ohara2021confinement}. Our work proposes a completely different strategy, which does not depend on the particular pinning energy landscape of the sample and does not require from additional geometrical confinement, enabling the nucleated structures to move freely over the whole continuous film. This constitutes an enormous advantage for the envisioned spintronic devices, as the information carriers can freely move in all plane directions. On top of that, our multilayer stacks have been tuned to have nucleation temperatures between $\approx$ 300-350 K, \textit{i.e.}, within the range of operating values of technological devices. Our method also allows to reduce the topological charge of the structures via OOP magnetic fields, as TSk ($2\pi$) can be effectively converted to skyrmions (1$\pi$) within the mT range. This constitutes an additional advantage of our method towards applications, as the topological charge has a direct influence on carrier velocities and/on read/write information process.\newline

Our engineered platform hosts TSk in continuous films of different orders, namely, 1, 2 and 3$\pi$. It remains an open question whether, at sufficiently high temperatures, higher order of TSk (5, 6, 7$\pi$ etc.) would become enabled, as already hypothesized by Gao \textit{et al.} \cite{gao2024aging}. Furthermore, our work enables the route for the experimental realization of hopfions \textit{in continuous films}, as the closure of the magnetization rotation in the $z$ dimension shall be forced by \textit{encapsulating} the multilayer in between to two stacks with higher PMA, mimicking the successful strategy proposed for TSk confined in nanodisks by N. Kent \textit{et al.} \cite{kent2021creation}.\newline


\begin{acknowledgments}
EMJ acknowledges the “Alexander von Humboldt Postdoctoral Fellowship”. Support from DFG (Priority Programme “Skyrmionics”), the ERC Synergy Grant 3D MAGiC (No. 856538) and Horizon Europe project NIMFEIA (101070290) are acknowledged.
\end{acknowledgments}


\section*{Author contributions statement}

E.M.J. conceived the experiments, methodology and prepared the samples; N.K. carried out the micromagnetic calculations; E.M.J, M.G.F. and J.M.M. conducted the experiments; E.M.J. and N.K. analyzed and discussed the results. M.K. supervised the project. All authors reviewed and contributed to the writing of the manuscript.

\section*{Competing interests}
The authors declare no competing interests.


\bibliography{references}


\end{document}


\preprint{AIP/123-QED}

\title{Supplemental Material to: Experimental realization of metastable target skyrmion states in continuous films}

\author{Elizabeth M. Jefremovas*}
\affiliation{Institute of Physics, Johannes Gutenberg University Mainz, Staudingerweg 7, 55128 Mainz, Germany}
\email{martinel@uni-mainz.de; **klaeui@uni-mainz.de}
\author{Noah Kent}%
\affiliation{Research Laboratory of Electronics, Massachusetts Institute of Technology, Cambridge, MA, United States of America}%
\author{Jorge Marqu\'es--March\'an}
\affiliation{Institute of Materials Sciences of Madrid -- CSIC, 28049 Madrid, Spain}
\author{Miriam G. Fischer}
\affiliation{Institute of Physics, Johannes Gutenberg University Mainz, Staudingerweg 7, 55128 Mainz, Germany}
\author{Agustina Asenjo}
\affiliation{Institute of Materials Sciences of Madrid -- CSIC, 28049 Madrid, Spain}
\author{Mathias Kl{\"a}ui**}
\affiliation{Institute of Physics, Johannes Gutenberg University Mainz, Staudingerweg 7, 55128 Mainz, Germany}
\email{klaeui@uni-mainz.de}

\date{\today}

\maketitle

\section{\label{continuous_film_n_2} Continuous film \textit{vs.} patterned rectangle}
\begin{figure*}
\includegraphics[scale=3]{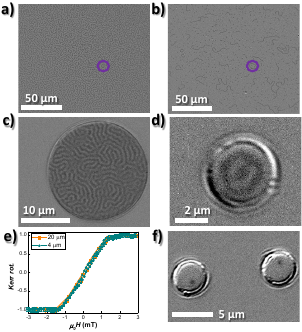}
\caption{\label{disquitos} \textbf{a)} and \textbf{b)} Kerr images corresponding to $n = $2 stack for the continuous film, at zero and 0.88 mT OOP field, respectively, measured at 292 K after heating up to 314 K.  Bright (dark) contrast corresponds to magnetization in $-z$ ($+z$) directions.\textbf{c)} and \textbf{d)} include representative images for 20 and 4 $\mu$m disks, measured at 0 OOP and 292K with no temperature heating protocol. As it can be seen, the multidomain configuration is not the energy minimum, but the TSk, when the structures are small enough. \textbf{e)} provides a direct comparison of the OOP hysteresis loops corresponding to the disks shown in \textbf{c)} and \textbf{d)}. No significant differences are observed in both structures. \textbf{e)} Kerr image of 4$\mu$m disks at OOP = 1.04 mT. As it can be seen, the TSk annihilate through the prevalence of the central skyrmion. }
\end{figure*}

Lateral confinement effects have been excluded for the studied 200 $\mu m$ x 200 $\mu m$ square by comparing its behaviour to the continuous film (non--patterned). As it can be seen in Fig.S~\ref{disquitos}\textbf{a}, $m$-$\pi$ states are nucleated after heating up to 314 K also in the continuous film. Measurements are taken at 0 field and room temperature (292 K). Fig.S~\ref{disquitos}\textbf{b} shows a representative image at 0.88 mT OOP magnetic field, highlighting the robustness of such states against the field. \newline

To obtain an estimation about the size from which the effect of lateral confinement becomes relevant, we have patterned a collection of disks, from 20 down to 1 $\mu$m diameter, with electron beam lithography, and studied the magnetic structure at 0 field and room temperature (\textit{i.e.}, without ramping temperature up). As it can be seen in Fig.S~\ref{disquitos}\textbf{c} and \textbf{d}, the geometric constraints are significant for diameters $\leq$ 4 $\mu$m, as a closed TSk structure is induced by the geometrical shape instead of the multidomain state. A comparison between the hysteresis loops of the 20 and 4 $\mu$m diameter disks is presented in Fig.~\ref{disquitos}\textbf{e)}, from where no significant differences are observed. The periodicity of the multidomain state is 1.6(2) $\mu$m, in good agreement to the continuous film \cite{eli}. The TSk nucleated in the 4 $\mu$m disk [see Fig.~\ref{disquitos}\textbf{d)}] comprises the circular domain + the inner skyrmion, being the width of each ring 210$\pm$ 20 nm. Therefore, it is expected that a TSk can be accommodated in disk sizes around 4 $\mu$m. Fig.~\ref{disquitos}\textbf{f)} illustrates the annihilation of the TSk through the inner skyrmion at an OOP field of 1.04 mT. \newline

\section{\label{Multidomain} Multidomain states at room temperature}
\begin{figure*}
\includegraphics[scale=0.65]{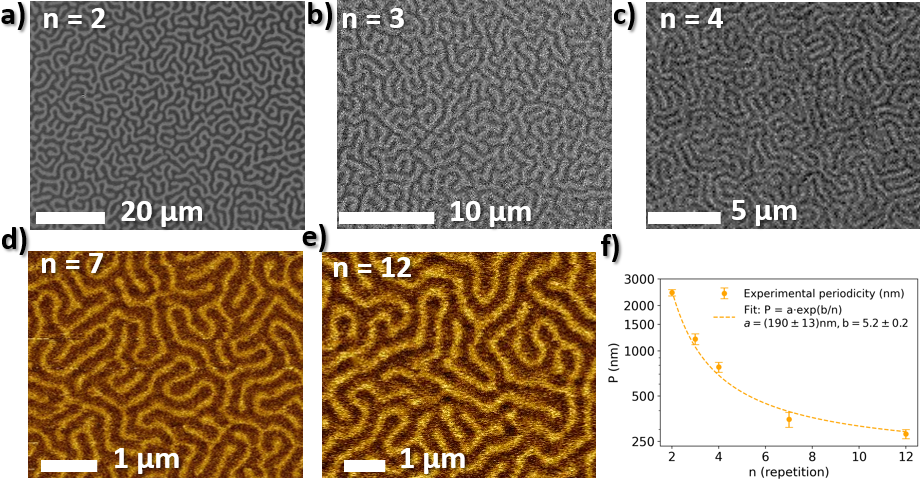}
\caption{\label{multidomain} \textbf{a)}--\textbf{e)} Multidomain pattern of $n =$ 2, 3, 4, 7 and 12 repetitions measured at 0 field after saturation at room temperature. In \textbf{a)}-\textbf{c)}, dark represents $+z$ direction, while in \textbf{d)} and \textbf{e)}, brown represents $+z$ magnetization. In \textbf{f)}, the periodicity lengths $P$ are fitted following a phenomenological $P = a \cdot exp [{\frac{b}{n}}]$ decay, being $a =$ 190$\pm$13 nm and $b =$ 5.2$\pm$0.2. Note the Y--axis is in log. scale.}
\end{figure*}

Fig.S~\ref{multidomain} includes the multidomain ground state of the $n =$2, 3, 4, 7 and 12 repetitions after saturation at room temperature ($T = $ 292 K). The corresponding periodicity lengths (calculated through 2D-FFT) are $P=$ 2440$\pm$120, 1390$\pm$70, 800$\pm$40, 320$\pm$20, and 280$\pm$20 nm, respectively, following the expected exponential decay behaviour with increasing dipolar coupling \cite{lemesh2017accurate, cape1971magnetic, kaplan1993domain, eli} (see fitting in Fig.S~\ref{multidomain}\textbf{f}). No topological texture is nucleated at RT without prior temperature increase. \newline

\section{\label{time} Target Skryrmions: Stability over time}
\begin{figure}
\includegraphics[scale=2.5]{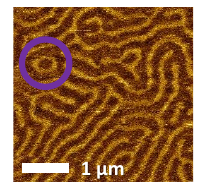}
\caption{\label{tiempo} MFM image at 0 field and $T =$ 292 K corresponding to $n =$ 7 repetition stack after 4 months. Brown (yellow) contrast corresponds to $+z$ ($-z$) magnetization. The nucleated TSk (encircled in purple) remained stable over time.}
\end{figure}

Fig.~\ref{tiempo} includes a MFM image measured at $T =$ 292 K and no external applied field corresponding to a TSk nucleated in $n =$ 7 after 4 months. As it can be seen, once nucleated, the structures remain stable over time and temperature. \newline


\bibliography{references}